\documentclass[a4paper]{cas-sc}
\usepackage{cite}
\usepackage{amsmath,amssymb,amsfonts}
\usepackage{algpseudocode}
\usepackage{graphicx}
\usepackage{algorithm}

\usepackage{textcomp}
\usepackage{xcolor}
\usepackage{xspace}
\usepackage{etoolbox}
\usepackage{hyperref}
\usepackage{booktabs}

\usepackage[numbers]{natbib}

\usepackage{graphicx}
\usepackage{url}
\def\tsc#1{\csdef{#1}{\textsc{\lowercase{#1}}\xspace}}
\tsc{WGM}
\tsc{QE}
\tsc{EP}
\tsc{PMS}
\tsc{BEC}
\tsc{DE}

\begin{document}
\let\WriteBookmarks\relax
\def\floatpagepagefraction{1}
\def\textpagefraction{.001}
\shorttitle{}
\shortauthors{Yujie Li et~al.}
\title[mode = title]{Continuous Multi-Task Pre-training for Malicious URL
Detection and Webpage Classification}

\tnotetext[1]{This work was supported in part by the National Natural Science Foundation of China under Grant 61902342, and in part by the Defense Industrial Technology Development Program under Grant JCKY 2021602B002.}

\author[1,4]{Yujie Li}
\cormark[0]
\ead{liyujie2003@bupt.edu.cn}

\credit{Conceptualization of this study, Methodology, Software}

\affiliation[1]{organization={The State Key Laboratory of Blockchain and Data Security, Zhejiang University},
                city={Hangzhou},
                postcode={330000}, 
                country={China}}

\author[3]{Yiwei Liu }
\cormark[0]
\ead{yiweiliu_disecc@163.com}

\author[3]{Peiyue Li }
\cormark[0]
\ead{lipeiyue@ppsuc.edu.cn}

\author[2,5]{Yifan Jia }
\cormark[0]
\ead{jiayf@hrbeu.edu.cn}

\author[4]{Yanbin Wang }[type=editor,auid=000,bioid=1,orcid=0000-0003-1682-5712]
\cormark[1]
\ead{11921107@zju.edu.cn}

\credit{Data curation, Writing - Original draft preparation}

\affiliation[2]{organization={Hangzhou Research Institute of Xidian University},
                postcode={310027}, 
                city={Hangzhou},
                country={China}}


\affiliation[3]{organization={Defence Industry Secrecy Examination and Certification Center},
                city={Beijing},
                postcode={065007}, 
                country={China}}

\affiliation[4]{organization={Shenzhen MSU-BIT University},
                city={Shenzhen},
                postcode={518172}, 
                country={China}}

\affiliation[5]{organization={Yantai Research Institute, Harbin Engineering University},
                city={Yantai},
                postcode={264000},
                country={China}}
                
\cortext[cor1]{Corresponding author}

\begin{abstract}
Malicious URL detection and webpage classification are critical tasks in cybersecurity and information management. In recent years, extensive research has explored using BERT or similar language models to replace traditional machine learning methods for detecting malicious URLs and classifying webpages. While previous studies show promising results, they often apply existing language models to these tasks without accounting for the inherent differences in domain data (e.g., URLs being loosely structured and semantically sparse compared to text), leaving room for performance improvement. Furthermore, current approaches focus on single tasks and have not been tested in multi-task scenarios.

To address these challenges, we propose {\scshape urlBERT}, a pre-trained URL encoder leveraging Transformer to encode foundational knowledge from billions of unlabeled URLs. To achieve it, we propose to use 5 unsupervised pretraining tasks to capture multi-level information of URL lexical, syntax, and semantics, and generate contrastive and adversarial representations. Furthermore, to avoid inter-pre-training competition and interference, we proposed a grouped sequential learning method to ensure effective training across multi-tasks. Finally, we leverage a two-stage fine-tuning approach to improve the training stability and efficiency of the task model. To assess the multitasking potential of {\scshape urlBERT}, we fine-tune the task model in both single-task and multi-task modes. The former creates a classification model for a single task, while the latter builds a classification model capable of handling multiple tasks. We evaluate URLBERT on three downstream tasks: phishing URL detection, advertising URL detection, and webpage classification. The results demonstrate that {\scshape urlBERT} outperforms standard pre-trained models, and its multi-task mode is capable of addressing the real-world demands of multitasking. The code is available at \href{https://github.com/Davidup1/URLBERT}{https://github.com/Davidup1/URLBERT}.

\end{abstract}

\maketitle

\section{Introduction}
Uniform Resource Locators (URLs) are serve as fundamental elements for navigating and comprehending the web. Malicious URLs Detection and Web Content Classification (such as web topic classification and advertising URL detection) are crucial tasks in critical domains such as cybersecurity, personalized online services, and web engineering. URLs have been used as important and easily accessible data sources for training malicious URL detection models or various webpage classification models, leveraging their inherent characteristics and the valuable semantic information they contain.

Previous studies developed isolated, domain-specific models that struggle to handle real-world multi-task scenarios\cite{sahoo2017malicious,baykan2009purely,sahingoz2019machine,opara2024look,ali2024efficient,sun2025ethereum}. These approaches were limited by fully supervised paradigms with several constraints:1)Reliance on domain expert engineering: Malicious URL detection required cybersecurity experts to extract features like suspicious patterns and threat indicators, while URL advertising detection needed marketing experts to focus on keywords and tracking parameters. 2) Redundant modeling overhead: Due to differences in subdomain data (e.g., between malicious URLs and advertising URLs), separate models needed to be trained for each subdomain, resulting in redundant training costs. 3)Performance validation dependent on labeled data: Most previous studies relied on supervised learning with labeled data, making model performance constrained by the quantity and quality of data labels. However, labeled data is often limited, and labels for malicious URLs typically contain significant noise. 

Recent advances in deep learning, particularly Transformer-based models\cite{tsai2024toward,do2022deep} such as BERT, offer promising alternatives to traditional methods. As a self-supervised learning framework, BERT is pre-trained on large-scale unlabeled corpora, which effectively alleviates the reliance on high-quality labeled data, a significant bottleneck in URL-related tasks. Furthermore, BERT automatically learns hierarchical and contextualized representations of input sequences, reducing or even eliminating the need for domain-specific feature engineering by experts. More importantly, as a flexible foundation model, BERT can be fine-tuned for various downstream applications, enabling a unified approach to handle different tasks such as phishing detection, advertisement classification, and topic categorization. These characteristics make BERT a natural candidate for building scalable and efficient URL understanding systems that overcome the limitations of traditional URL classification pipelines. However, direct application of BERT to URL-related problems is suboptimal due to structural and semantic disparities between natural language text and URLs. Key challenges include: 1) The simple yet loose structure of URLs follows a basic standard, and their components (such as paths and query parameters) can be freely added or modified without following strict patterns. 2) URLs have sparser semantics. 3) The high level of noise, including irrelevant or disruptive elements like session identifiers and cache-busting parameters. These challenges necessitate a dedicated URL representation model that captures structural, sequential, and semantic patterns from large-scale unlabeled URL data.

In this paper, we introduce \textsc{urlBERT}, a pre-trained URL encoder that quickly builds models for malicious URL detection and web content classification with simple fine-tuning. \textsc{urlBERT} learns foundational URL information from a corpus of 3 billion URLs through unsupervised pre-training, enabling cross-task knowledge sharing and eliminating the need for manual feature engineering and repetitive modeling. We address pre-training challenges with five tasks, including self-supervised contrastive learning (SCL) and virtual adversarial training (VAT) to overcome noise and learn robust representations, masked language modeling (MLM) to capture URL semantics, and two token detection tasks—replaced token detection (RTD) and shuffled token detection (STD)—to learn structural and sequential features from the loose URL structure. Additionally, we propose a grouped continual learning method to avoid conflicts between tasks. Finally, we use a two-stage fine-tuning strategy to improve model stability and reduce resource consumption. Our results in both single-task and multi-task scenarios demonstrate the advantages of our model over previous approaches.

In contrast to existing approaches that either apply general-purpose language models to URL tasks or focus on single-task fine-tuning, our proposed \textsc{urlBERT} introduces two key innovations. First, we design a set of self-supervised pre-training tasks specifically tailored to the structural and semantic characteristics of URLs. These tasks allow the model to learn robust representations from raw URLs without relying on textual syntax or domain-specific assumptions. Second, \textsc{urlBERT} supports multi-task learning through a unified encoder, enabling a single model to effectively handle diverse downstream applications such as phishing detection, advertising classification, and topic categorization. This combination of URL-specific representation learning and multi-task generalization allows \textsc{urlBERT} to outperform prior models that are either narrowly focused or not optimized for the unique challenges of URL data.

The remainder of this paper is organized as follows:
Section \ref{section 2} surveys the related work in URL classification and pre-trained models.
Section \ref{section 3} introduces the architecture, tokenizer design, and pre-training tasks of \textsc{urlBERT}.
Section \ref{section 4} describes the experimental setup, datasets, and evaluation protocols.
Section \ref{section 5} presents empirical results and comparisons.
Section \ref{section 6} discusses key findings and future research directions.
Finally, Section \ref{section 7} concludes the work.

Our primary contributions are as follows:
\begin{itemize}
    \item We propose \textsc{urlBERT}, to our knowledge, the first pre-trained URL-Transformer encoder capable of performing various malicious URL detection and webpage classification tasks.
\item We propose to learn  multi-level information of URLs through five pre-training tasks, including their structure, sequence, semantics, and contrastive and adversarial features.
    \item We propose a grouped sequential learning method to address the training conflicts among the five pre-training tasks.
    \item We leverage a two-stage fine-tuning to improve model stability and reduce training resource consumption. 
    \item We create task models for both single-task and multi-task settings and empirically validate that our pre-training method achieves superior results compared to previous approaches in both cases.
\end{itemize} 


 




\section{Related Work}\label{section 2}
We review related work from two perspectives: the tasks of URL detection or classification, and the technique of pre-trained models.

\subsection{URL Detection or Classification}

URLs transmit various types of internet content, and URL detection and classification are crucial for internet security and online business operations. Common tasks include malicious URL detection, webpage topic classification, and advertising URL detection. Early approaches used predefined rule lists for URL classification. However, challenges with timeliness and scalability led researchers to adopt machine learning algorithms, training detection or classification models using task-specific labeled datasets\cite{dmoz1,grambeddings,Mendeley} .

These algorithms can be broadly categorized into two approaches: shallow model-based and deep model-based methods. Shallow model approaches utilize algorithms like Random Forests and Support Vector Machines with fewer hyperparameters, but typically require domain expert-engineered features to enhance model capabilities.  Deep model-based approaches leverage neural networks' powerful nonlinear learning capabilities to learn better representations directly from raw URLs without additional expert input. These methods employ convolutional neural networks, recurrent neural networks, and attention networks\cite{urlnet,ozcan2023hybrid,wei2020accurate,srinivasan2021durld,tian2025past}. Furthermore, in current deep models, the initial step of converting URLs into numerical format is crucial, accomplished either through one-hot encoding or using Natural Language Processing (NLP) techniques like Word2Vec\cite{mikolov2013distributed}, TextCNN\cite{kim2014convolutional}, and FastText\cite{joulin2016bag}.

While URL detection and classification have advanced through machine learning developments, current methods are typically tailored to specific tasks and heavily rely on well-defined labeled datasets, resulting in numerous limitations and resource consumption. 

\subsection{Pretrained Models}
Recent advancements in pre-trained models have raised hopes of overcoming the limitations of traditional techniques. The Transformer-based Bidirectional Encoder Representations (BERT) \cite{devlin2018bert} introduced the "pre-training-fine-tuning" paradigm, which has revolutionized machine learning, showcasing the powerful capabilities of Transformers and the value of large amounts of unlabeled data. Several studies have attempted to adapt general-purpose language models to URL-related tasks. For example, \cite{su2023bert, elsadig2022intelligent, he2023method, seyyar2022detection, liu2025pmanet,liu2024transurl} explore using BERT or RoBERTa as feature extractors for phishing or malicious URL detection. In these approaches, the model is pre-trained on natural language corpora and then applied to URLs without any domain adaptation. As a result, performance is often limited by the semantic and structural mismatch between URL strings and natural language.

To address these issues, some efforts \cite{otieno2023detecting, liu2023malicious, artene2021using} introduce domain-specific tokenization or fine-tuning strategies. For instance, \cite{haynes2021lightweight} builds a custom vocabulary for phishing URL classification using BERT, while \cite{artene2021using} fine-tune BERT models with labeled phishing datasets. Nevertheless, these methods remain constrained by small-scale task-specific supervision and do not explore unsupervised pre-training on large-scale raw URL corpora. Recently, \cite{wang2023large} introduced phishBERT, a model specifically designed for phishing URL detection, demonstrating state-of-the-art results. Yet, it still focuses on a single task and lacks a pre-training method tailored to URL data characteristics. 

In contrast, our approach, \textsc{urlBERT}, introduces a URL-native pre-training framework that incorporates five self-supervised learning objectives, each carefully designed to capture the structural irregularities, sequential patterns, and semantic sparsity unique to URLs. Unlike prior models that either reuse language-based pre-training or focus on narrow tasks, URLBERT learns robust representations directly from massive unlabeled URL corpora. Furthermore, \textsc{urlBERT} is designed as a general-purpose encoder, capable of supporting multiple downstream tasks. This makes it highly adaptable in both single-task and multi-task settings. More importantly, as a domain-specific foundational model, \textsc{urlBERT} can serve as a versatile backbone for other URL-related frameworks\cite{liu2025pmanet}, facilitating the development of custom detection pipelines or plug-in modules in broader security and content analysis systems.

\section{Methods}\label{section 3}
Our \textsc{urlBERT} adopts an architecture and parameter set identical to BERT-base, featuring a 12-layer Transformer network with 12 attention heads, 768 hidden units, and a maximum input length of 512 tokens. For a detailed description, please refer to the original BERT paper by Devlin et al. {\cite{devlin2018bert}}. Consequently, we will focus our discussion on the core technical aspects of \textsc{urlBERT}, including URL tokenization, pre-training objectives, and fine-tuning methods.

\begin{figure*}[h]
    \centering
    \includegraphics[width=.6\linewidth]{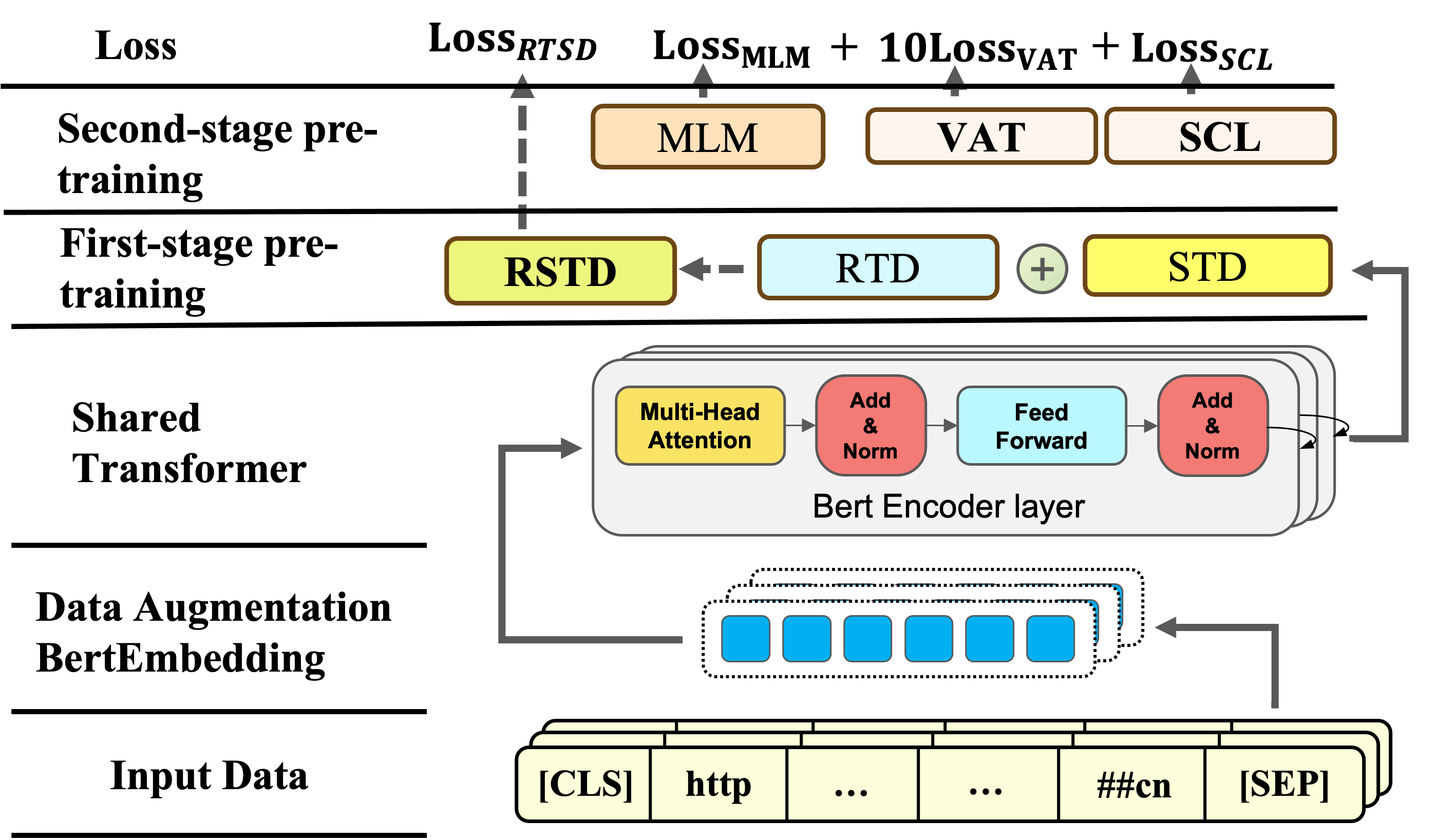}
    \caption{The pre-training task framework of \textsc{urlBERT}. In the first phase, the model is trained on the RSTD task, which involves detecting shuffled and replaced tokens. In the second phase, training targets Masked Language Model, virtual adversarial training, and contrastive learning for pre-training tasks.}
    \label{fig:Framework}
\end{figure*}

\subsection{URL-tokenizer}
Despite pre-trained models with versatile text tokenizers trained on extensive internet data, domain-specific tokenization is crucial for URLs. URLs differ from standard text, containing protocol identifiers, domain names, path components, and query parameters—each carrying unique information. A dedicated URL tokenizer effectively dissects these elements, enabling precise feature extraction. 

For simplicity, we retrain the tokenizer using BERT's tokenization algorithm on our collected URL corpus. Initially, we set a vocabulary size of 30,000 tokens, which the tokenizer refines during training to prioritize frequent and relevant subwords. To ensure inputs are normalized and context-focused rather than stylistically diverse, our tokenizer uses a case-insensitive approach, converting all letters to lowercase. 

In revising our token strategy to better align with the specific needs of our model, we have made deliberate choices about which BERT tokens to retain and which to omit. Unlike BERT, we omit the [SEP] token since our tasks do not involve predicting subsequent sentences. However, we retain the crucial [CLS] token for sequence classification tasks, utilizing its final hidden state as a key aggregate representation. The [PAD] token ensures uniform sequence lengths within batches, essential for consistency. Within the MLM framework for pre-training, the [MASK] token remains pivotal, enhancing contextual comprehension by prompting the model to predict masked tokens. Additionally, the [UNK] token represents out-of-vocabulary words, maintaining tokenization integrity. We also introduce two new special tokens: [REP] for token replacement tasks and [SHU] for predicting shuffled tokens during specific pre-training exercises.

\subsection{Pre-training Tasks}
In order to effectively capture URL information, we apply five distinct pre-training tasks. These tasks assist in learning the structure, sequence, and semantics of URLs, and produce robust, discriminative representations that are resistant to noise. The training process is depicted in Figure \ref{fig:Framework}.

\subsubsection{\textbf{Task 1: Replaced Token Detection (RTD)}}
The structure of URLs is simple yet loose, which leads to several challenges such as inconsistency in patterns, the flexibility of URL components (like paths and query parameters). This variability makes it difficult for models to reliably capture structural features and interpret the inherent meaning of URLs. To address this, we learn the structural information of URLs using the replaced Token Detection (RTD) task. In this task, a portion of the input URL is randomly replaced with tokens from the URL vocabulary. The goal is for the model to learn to identify the replaced tokens within the URL. By effectively distinguishing between original and replaced tokens in the URL structure, the model becomes better at understanding the roles and relationships of different URL components, such as protocol, domain, path, and query parameters. This approach improves the model's ability to capture the underlying structural characteristics of URLs, making it more robust in tasks like malicious URL detection and webpage classification.

\subsubsection{\textbf{Task2: Shuffled Token Detection (STD)}}  
The sequence of tokens in a URL reflects the hierarchical structure of the web resource being accessed, including the domain, subdomain, path, and query parameters. This hierarchical order is crucial for understanding the relationships between different components of the URL and the content or service it points to. For instance, the path segment often indicates a specific resource or page, while query parameters may specify additional filtering or sorting options for that resource. To capture the sequential information of URLs, we use the Shuffled Token Detection (STD) task. This task involves randomly shuffling portions of the input URL token sequence and training the model to determine whether each token is in the correct position.  The STD task enhances the model's understanding of how URL tokens interact in a specific order to form meaningful and functional web addresses.

\subsubsection{\textbf{Task3: Masked Language Model (MLM)}}
Masked Language Models (MLMs) learn context-dependent semantics by pre-training on masked tokens, which allows them to capture deep syntactic and semantic patterns within URL. By masking random tokens during training, these models are trained to predict the missing tokens based on their surrounding context, enabling them to develop a rich understanding of token dependencies and the underlying relationships between entities. This process helps the model generate more coherent and contextually relevant predictions when applied to downstream tasks. Note that we are not using MLM in isolation; rather, we integrate it with Virtual Adversarial Training (VAT) tasks in a collaborative manner. For more detailed information on this integration, please refer to section \ref{MLM_VAT}.

\subsubsection{\textbf{Task4: Self-supervised Contrastive learning (SCL)}}
The self-supervised Contrastive Learning (SCL) task aims to distinguish subtle differences between URL and derive distinguishable representations. It maximizes consistency among various transformations (positive samples) of the same instance while minimizing consistency between different instances (negative samples). 

In order to implement SCL, we propose a data augmentation method by combining Dropout and adversarial learning techniques to generate positive and negative samples required for SCL training.

\begin{equation}
\hat{x} \leftarrow \frac{1}{1-p} \cdot x \cdot \text{MASK}(p)
\end{equation}

Here, $p$ denotes the probability of neuron extraction, and $MASK$ is a tensor with the same shape as the sample $x$, where each element is 1 with probability $p$ and 0 otherwise. 

Next, we employ the Fast Gradient Sign Method (FGSM)\cite{goodfellow2014explaining} for adversarial learning. FGSM adjusts data samples by leveraging gradient descent during training to introduce perturbations, aimed at maximizing the model's loss and disrupting its convergence. The FGSM attack algorithm is as follows:


\begin{equation}
\hat{x} \leftarrow x + \alpha \cdot \text{sign}(\nabla g(x,y \mid \theta))
\end{equation}

Here, $x$ is the original token sample, and $\hat{x}$ is the augmented sample generated. $\theta$ refers to the inference process of the sample in the MLM task within the BERT model. $sign(\bullet )$ regularizes the gradients, and $\alpha$ indicates the strength of the added regularization perturbation.

Inspired by ConSERT\cite{yan2021consert} and SimCSE\cite{gao2021simcse}, the SCL task in {\scshape urlBERT} incorporates token-level data augmentation to create positive and negative samples. Unlike ConSERT, {\scshape urlBERT} uses BERT's Embedding layer dropout module to generate two augmented samples per input. Subsequently, the FGSM\cite{goodfellow2014explaining} is applied to each sample independently to create adversarial samples, thereby enhancing the effectiveness of data augmentation.

A batch of $N$ token embeddings produces $2N$ augmented data samples. Each batch includes one positive sample and $2(N-1)$ negative samples. The objective of the SCL task is to minimize the distance between positive samples and maximize the distance between negative samples. {\scshape urlBERT} employs the Normalized Temperature-scaled Cross-Entropy Loss (NT-Xent):


\begin{equation}
\label{eq:example}
\mathcal{L}_{\mathrm{i}, \mathrm{j}}=-\log \frac{\exp \left(\operatorname{sim}\left(\mathrm{r}_{\mathrm{i}}, \mathrm{r}_{\mathrm{j}}\right) / \tau\right)}{\sum_{\mathrm{k}=1}^{2\mathrm{~N}} 1_{[\mathrm{k} \neq \mathrm{i}]} \exp \left(\operatorname{sim}\left(\mathrm{r}_{\mathrm{i}}, \mathrm{r}_{\mathrm{k}}\right) / \tau\right)}
\end{equation}

Here, $r_i$ and $r_j$ are representations obtained from the final layer of the BERT encoder after encoding augmented data samples. The function $sim(\bullet)$ denotes cosine similarity, and $\tau$ is the temperature coefficient. After computing this loss for each data point and averaging, it yields the loss for this task, denoted as ${Loss}_{con}$. The pseudocode for the SCL task is depicted in Algorithm \ref{alg: alg1}.


\begin{algorithm}[t]
    \caption{SCL Task}
    \textbf{Input:} \(\chi\): dataset, \(N\): data batch, \(\theta\): shared {\scshape urlBERT} model, \(\tau\): learning rate\\
    \textbf{Output:} \(\theta\)
    \label{alg: alg1}
    \begin{algorithmic}[1]
        \For {\( N\in\chi\)}
        \State \(2N\leftarrow \text { data augmentation }(N, \theta)\)
        \For {\( X\in2N\)}
        \State \(y \leftarrow \text{model encoding}(x, \theta)\)
        \State \({Loss}_{i} \leftarrow \text{calculate NT-Xent}(y, x)\)
        \EndFor
        \State \({Loss}_{con} \leftarrow \frac{1}{2N} \sum_{1}^{2N} {Loss}_{i}\)
        \State \({g}_{con} \leftarrow \nabla {Loss}_{con}\)
        \State \(\theta \leftarrow \theta-\tau {g}_{con}\)
        \EndFor
        \State
        \Return \(\theta\)
    \end{algorithmic}
\end{algorithm}

\subsubsection{\textbf{Task5: Virtual Adversarial Training}}
\label{MLM_VAT}
URLs often exhibit high levels of noise due to the presence of non-informative or redundant characters (such as query parameters or session IDs). These factors can make it difficult for machine learning models to effectively capture meaningful patterns in URLs. As a result, models trained solely on raw URLs may struggle with generalization and fail to accurately process or predict patterns within new or unseen URLs.

Adversarial training techniques can help mitigate this issue by increasing the model's robustness to noise and perturbations. By introducing adversarial training, model is forced to learn more invariant and generalizable features. This enables the model to focus on the most critical aspects of the URL, despite the noise and inconsistencies in the input data. 

We implement VAT task in {\scshape urlBERT} by introducing regularized perturbations in the token embedding space to generate adversarial samples corresponding to unlabeled URL data. We continue using FGSM for data augmentation, but adopt KL divergence as the new objective function to measure discrepancies between original and augmented sample outputs The formula of the loss function is as follows: 
\begin{equation}
\label{eq:loss_div}
loss_{div} = \mathrm{D}(p \| q) = \sum_{i=1}^{N} p(y_{i}) \cdot (\log(p(y_{i})) - \log(q(x_{i})))
\end{equation}
Herein, $p(y)$ represents the data distribution of model outputs for augmented samples,  $q(x)$ denotes the model output for original samples. During data augmentation, the gradients of input data are computed with $KL$ divergence, and small perturbations, subject to a $l2$ norm constraint, are added in the direction of gradient ascent to generate augmented samples.
\begin{equation}
\label{eq:update_rule}
\hat{x} \leftarrow x + \alpha \, \text{sign}(\nabla g(x \mid \theta, loss_{KL}))
\end{equation}

\begin{algorithm}[t]
    \caption{VAT Task}
    \textbf{Input:} \(\chi\): dataset, \(x\): data batch, \(\theta\): shared {\scshape urlBERT} model, \(\tau\): learning rate, \(\sigma ^ 2 \): the variance of the noise, \(\alpha\): weight of VAT task loss, \(\Pi\): the normalization operation, \(\delta \): introduced disturbance, \(\mu\): step size of perturbation\\
    \textbf{Output:} \(\theta\)
    \label{alg: alg2}
    \begin{algorithmic}[1]
        \For{\(x\in\chi\)}
            \State \(\delta \sim N\left(0, \sigma^{2}\right)\)
            \State \(y_{1} \leftarrow \text { model encoding }(\theta, x+\delta)\)
            \State \(y_{2} \leftarrow \text {model encoding }(\theta, x)\)
            \State \({Loss}_{adv} \leftarrow \text { calculate } loss_{KL}\left(y_{1}, y_{2}\right)\)
            \State \(g_{adv} \leftarrow \nabla \operatorname{Loss}_{adv}\)
            \State \(\delta \leftarrow \Pi\left(\delta+\mu g_{adv}\right)\)
            \State \(y_{3} \leftarrow \text{model encoding }(\theta, x+\delta)\)
            \State \({Loss}_{V A T} \leftarrow \text { calculate } loss_{KL}\left(y_{3}, y_{2}\right)\)
            \State \(g_{\theta} \leftarrow \alpha \nabla {Loss}_{V A T}\)
            \State \(\theta \leftarrow \theta-\tau g_{\theta}\)
        \EndFor
        \State
        \Return \(\theta\)
    \end{algorithmic}
\end{algorithm}

{\scshape urlBERT} uses gradient forward propagation, initializing perturbations with 0 mean and 1 variance. The BERT encoder processes samples to obtain logits and ({logits}{adv}). The virtual adversarial loss (({Loss}{VAT})) is computed using KL divergence ($loss_{KL}$). A regularization perturbation is then added in the reverse gradient direction to maximize ({Loss}{VAT}). While this process can be iterated, {\scshape urlBERT} applies perturbation once. The final loss combines weighted ({Loss}{VAT}) and MLM task loss. Algorithm \ref{alg: alg2} details the VAT task training process.

\subsection{Grouped Sequential Learning}
To capture the intricate linguistic features of URLs, \textsc{urlBERT} is trained across a diverse array of tasks to deeply grasp URL syntax, structure, and semantics. However, training these tasks together may lead to challenges like task interference. Additionally, our pre-training involves advanced token manipulations—replacements, shuffling, and more—beyond basic masking, posing further challenges in learning clarity, and optimization.

To mitigate such interference, we propose Grouped Sequential Learning (GSL), a lightweight but effective strategy that organizes pre-training tasks into semantically coherent groups and trains them in a staged manner. Specifically, we divide our five pre-training tasks into two stages where each builds on parameters from the previous one.

Stage 1 includes structure-oriented tasks, Replaced Token Detection (RTD) and Shuffled Token Detection (STD), which focus on learning URL structure and sequence coherence. These tasks are jointly performed using a three-way token-level classification to determine whether a token is shuffled, replaced, or original. Specifically, we substitute 5\% of tokens with shuffled versions and an additional 5\% with random tokens in each sample. The loss function employed for this task is cross-entropy loss, averaged over all input tokens:

\begin{equation}
    \mathcal{L}_{\text {RSTD}}=-\frac{1}{N} \sum_{i=1}^{N} \sum_{j=1}^{3} y_{i j} \log p_{i j}\left(x_{i}\right)
\end{equation}

where  $p_{i j}\left(x_{i}\right)$ represents the probability of the  $i --th$ input token  $x_{i}$  predicted as shuffled  ($j=1$) , randomized  ($j=2$) , or original  ($j=3$)  by a model.  $y_{i j}$  is the corresponding target label.

Stage 2 includes semantics-oriented tasks, Masked Language Modeling (MLM), Self-supervised Contrastive Learning (SCL), and Virtual Adversarial Training (VAT), which encourage semantic understanding and robustness under perturbation. These tasks are weightedly summed in the training loss:
\begin{equation}
\operatorname{Loss} = \operatorname{Loss}_{con} + \operatorname{Loss}_{MLM} + 10 \operatorname{Loss}_{VAT}
\end{equation}

Our continual learning approach draws inspiration from ERNIE 2.0, a multi-task pre-trained model. For details on continual learning, please refer to \cite{sun2020ernie}. Unlike ERNIE 2.0, where tasks are sequentially learned, our method groups tasks together for joint training within each phase of continual learning. Each stage initializes from the parameters learned in the previous one. This continual pre-training strategy ensures that foundational knowledge (e.g., structure patterns) is well established before optimizing for deeper semantic and discriminative features.

While our method is superficially similar to curriculum learning,which trains from easier to harder tasks, and multi-task scheduling, which dynamically adjusts task weights or orders, it differs in two key aspects: 1) GSL groups tasks based on their learning focus (i.e., structure and semantics), rather than assumed difficulty, which is often ambiguous or data-dependent in unsupervised settings; 2) Unlike multi-task schedulers that interleave or reweight tasks within the same epoch, GSL performs stage-wise full optimization of grouped tasks, reducing gradient conflict and improving stability.

\subsection{Fine-tuning}
The deep nested structure and vast number of parameters in pre-trained models present a dilemma when fine-tuning the weights of \textsc{urlBERT} models directly. If the learning rate is too high, it may disrupt the knowledge system acquired by the pre-trained model from massive data during pre-training. Conversely, if the learning rate is too low, it may impede the convergence of the pre-trained model during task-specific learning.
Inspired by \cite{wang2019tune}, we use a two-stage fine-tuning strategy. In the first stage, the parameters of the pre-trained model remain fixed, and only the upper-layer model added for specific tasks (in our case, a CNN followed by fully connected layers) is trained. In the second stage, we fine-tune the upper-layer model together with the pre-trained language model.

We employ different learning rates for these two phases. For the first stage, we experimented with learning rates of $10^{-2}$, $5 \times 10^{-3}$, $10^{-3}$, and $5 \times 10^{-4}$. In the latter stage, we set the learning rate to a smaller value, such as $10^{-3}$, $10^{-4}$, $5 \times 10^{-5}$, or $10^{-5}$. Our experiments revealed that optimal results are achieved with a learning rate of $2 \times 10^{-3}$ in the stage of training only the upper model, and $2 \times 10^{-5}$ in the later stage.

\subsection{Multi-Task Learning in Fine-tuning}
\label{MT}
Conventional fine-tuning of pre-trained models typically involves adding a top-layer model for supervised learning on specific tasks, generating a task-specific model for each task. In contrast, multi-task learning involves jointly optimizing for multiple tasks\cite{zhang2021multi}. It also is recognized as a complementary technique to language model pre-training, potentially improving text representation learning when combined.
In our study, we apply multi-task supervised learning during the URL classification task learning phase, aiming to leverage cross-task data to enhance performance in downstream tasks. The core of multi-task learning lies in configuring a task-specific layer for each task (in our case, classification layers for our three downstream tasks), with these task layers sharing the \textsc {urlBERT} encoder. 

\begin{algorithm}[t]
    \caption{Training a urlBERT-MT model.}
    \textbf{Input:} \(\Theta\): model parameters, \(T\): number of tasks, \(epoch_{max}\): maximum number of epochs\\
    \textbf{Output:} Trained model parameters \(\Theta\)
    \begin{algorithmic}[1]
        \State Initialize model parameters \(\Theta\) randomly.
        \State Pre-train the shared layers (i.e., the lexicon encoder and the transformer encoder).
        \State Set the max number of epoch: \(epoch_{max}\).
        \State \textit{//Prepare the data for T tasks.}
        \For{\(t\) in \(1, 2, ..., T\)}
            \State Pack the dataset \(t\) into mini-batch: \(D_t\).
        \EndFor
        \For{\(epoch\) in \(1, 2, ..., epoch_{max}\)}
            \State 1. Merge all the datasets: \(D = D_1 \cup D_2... \cup D_T\)
            \State 2. Shuffle \(D\)
            \For{\(b_t\) in \(D\)}
                \State \textit{//$b_t$ is a mini-batch of task t.}
                \State 3. Compute loss : \(L(\Theta)\)
                \State 4. Compute gradient: \(\nabla(\Theta)\)
                \State 5. Update model: \(\Theta = \Theta - \epsilon\nabla(\Theta)\)
            \EndFor
        \EndFor
        \State \Return \(\Theta\)
    \end{algorithmic}
\end{algorithm}

We employ a minibatch-based AdamW optimizer to learn the parameters of our model (i.e., the parameters of all shared layers and task-specific layers) as outlined in Algorithm 3. In each epoch, a mini-batch $b_t$ is selected (e.g., from among all 3 URL classification tasks), and the model is updated according to the task-specific objective for task $t$. This approach approximately optimizes the sum of all multi-task objectives. For classification tasks , we utilize the cross-entropy loss as the objective: 

\begin{equation}
-\sum_{c} \mathbb{1}(X, c) \log \left(P_{r}(c \mid X)\right)
\end{equation}
where  $\mathbb{1}(X, c)$  is the binary indicator  $ (0  or 1 ) $ if class label $c $is the correct classification for $X$. $P_{r}$ is softmax function.

\section{Experimental configuration}\label{section 4}

\begin{table*}[t!]
  \centering
  \caption{Details of \textsc{urlBERT} Corpus and Task-Specific Datasets}
  \resizebox{\textwidth}{!}{
  \begin{tabular}{|c|l|r|l|}
    \hline
    \multicolumn{4}{|c|}{\textbf{\textsc{urlBERT} Corpus}} \\
    \hline
    \textbf{URL Corpus} & \textbf{Description} & \textbf{URL Count} & \textbf{Source} \\
    \hline
    \(D^{1}\) & Web page URLs from Common Crawl & 2.85 billion & \href{https://commoncrawl.org/}{Common Crawl} \\
    \(D^{2}\) & URLs from top 10 million websites on Open PageRank & 338.71 million & \href{https://openpagerank.com/}{Open PageRank} \\
    \(D^{3}\) & URLs associated with expired domains & 538.2 million & \href{https://www.expireddomains.net/}{Expired Domains} \\
    \(D^{4}\) & URLs from various malicious sources & 8.26 million & Multiple Sources \\
    \(D^{5}\) & URLs fetched from the Open Directory Project & 3.86 million & \href{https://dmoz-odp.org/}{ODP} \\
    \hline
    \multicolumn{4}{|c|}{\textbf{Task-Specific Datasets}} \\
    \hline
    \textbf{Task} & \multicolumn{2}{l|}{\textbf{Dataset Details}} & \textbf{Sample Counts} \\
    \hline
    Phishing URL Detection & 
    \multicolumn{2}{l|}{Combined GramBedding and Mendeley datasets. Removed URLs $>$ 150} & 
    Phishing: 126,733 \\
    & \multicolumn{2}{l|}{characters and similar samples (similarity $>$ 0.7).} & 
    Benign: 237,889 \\
    \hline
    Webpage Topic Classification & 
    \multicolumn{2}{l|}{DmoZ URL Dataset, five categories: Games, Health, Kids,} & 
    Games: 56,477 \\
    & \multicolumn{2}{l|}{Reference, and Shopping.} & 
    Health: 60,097 \\
    & \multicolumn{2}{l|}{} & 
    Kids: 46,182 \\
    & \multicolumn{2}{l|}{} & 
    Reference: 58,247 \\
    & \multicolumn{2}{l|}{} & 
    Shopping: 95,270 \\
    \hline
    Advertising URL Detection & 
    \multicolumn{2}{l|}{Crawled from top 10,000 websites.} & 
    Advertising: 23,530 \\
    & \multicolumn{2}{l|}{} & 
    Non-advertising: 65,263 \\
    \hline
  \end{tabular}}
  \label{tab:urlbert_and_task_datasets}
\end{table*}

\subsection{Pretraining Corpus}
We curate a large-scale and highly diversified unlabeled URL corpus for pre-training {\scshape urlBERT}. This corpus contains 3.28 billion unique URLs, which we denote as
$D = unique( {D_{a}^{1}\cup D_{a}^{2}\cup...\cup D_ {a}^{n} })$, where each $D^{i}$ is a dataset acquired from distinct data sources. $D^{1}$ comprises 2.85 billion web page URLs, procured from \texttt{Common Crawl}\footnote{https://commoncrawl.org/}, an open-source initiative that traverses all publicly accessible URLs on the World Wide Web. $D^{2}$ encompasses a total of 338.71 million URLs obtained by crawling the top 10 million websites ranked on the \texttt{Open PageRank} platform. $D^{3}$ is a dataset\footnote{https://www.expireddomains.net/} containing 538.2 million URLs, predominantly associated with expired domains pertaining to activities such as pornography, gambling, scams, phishing, and other malicious activities. $D^4$ consists of 8.26 million URLs extracted from various malicious URL data sources, including \texttt{PhishTank}\footnote{https://phishtank.org/}, \texttt{Phishing.Database}\footnote{https://github.com/ mitchellkrogza/Phishing.Database}, \texttt{PhishRepo}, \texttt{PhishStorm}\footnote{https://research.aalto.fi/files/16859732/urlset.csv.zip}, \texttt{Iscx-url2016} \footnote{https://www.unb.ca/cic/datasets/url- 2016.html}, \texttt{Urlhaus}\footnote{https://urlhaus.abuse.ch}, \texttt{malware-host}\footnote{https://github.com/RamSet/ad-hosts-blocking/ blob/
main/malware-hosts}, \texttt{UT1}\footnote{ https://http://dsi.ut-capitole.fr/blacklists/}, \texttt{kaggle1}\footnote{ https://www.kaggle .com/datasets/taruntiwarihp/phishing-site-urls/code}, \texttt{kaggle2}\footnote{https://www.kaggle.com/datasets/sid321axn/malicious-urls-dataset}. Lastly, $D^5$ comprises 3.86 million URLs fetched from the \texttt{Open Directory Project (ODP)}\footnote{https://www.dmoz-odp.org/}. Taken together, $D$ covers over 3 billion URLs, which is a near-complete representation of all URLs on the Internet. 

The training corpus data is summarized in Table \ref{tab:urlbert_and_task_datasets}. Additionally, Figure \ref{hig} illustrates the distribution of URL lengths across both the training corpus and the three task-specific datasets. As evident from the figure, the URL length distribution in the training corpus comprehensively encompasses the statistical range of URL lengths observed in all task datasets.

\subsection{Description of Tasks and Datasets}
Our experiments focus on three downstream tasks: (1) Phishing URL Detection: Identifying URLs designed to deceive users into revealing sensitive information; (2) Webpage Topic Classification: Categorizing URLs into different webpage topics; and (3) Advertising URL Detection: Distinguishing advertisement URLs from non-advertisement URLs.

For these tasks, we utilize the following datasets:
\begin{enumerate}
\item For phishing URL detection, we combined two public datasets: GramBedding and Mendeley. To assess phishing detection performance more accurately, we first removed malicious URL samples with lengths exceeding 150 characters, as URL length is a significant distinguishing factor, and such lengthy malicious samples are easily identifiable. We then calculated the similarity between malicious URLs and eliminated samples with a similarity score greater than 0.7. Ultimately, our refined dataset comprised 126,733 phishing samples and 237,889 benign samples.
\item For Webpage Topic Classification, we employ the DmoZ URL Dataset, focusing on five categories: Games (56,477 URLs), Health (60,097 URLs), Kids (46,182 URLs), Reference (58,247 URLs), and Shopping (95,270 URLs).
\item For Advertising URL Detection, we crawl the top 10000 websites to gather 23,530 advertising URLs and 65,263 non-advertising URLs.
\end{enumerate}

Table \ref{tab:urlbert_and_task_datasets} provides a comprehensive overview of the details for each task dataset. All three task datasets exhibit class imbalance, intentionally designed to closely reflect real-world data distributions\cite{lopez2013insight}. This imbalance presents a significant challenge for classifier learning, mirroring the complexities encountered in practical applications\cite{fernandez2018learning,garcia2018dynamic}.

\section{Experiments}\label{section 5}
We evaluated the proposed URL encoder, \textsc{urlBERT}, on three downstream tasks: phishing URL detection, advertising URL detection, and webpage classification. Through experiments, we answer the following questions:

\begin{description}
    \item[Q1:] Can \textsc{urlBERT} serve as an effective static feature extractor for various URL-related tasks?
    \item[Q2:] How does \textsc{urlBERT} compare to existing pre-trained models and classical deep learning methods in terms of performance?
    \item[Q3:] What is the sensitivity of \textsc{urlBERT} to the scale of training data?
    \item[Q4:] How well does \textsc{urlBERT} perform in multi-task label learning scenarios?
\end{description}

\subsection{Baseline}
For comparison with our method, we select three classic pre-trained models (BERT\cite{devlin2018bert}, RoBERTa\cite{liu2019roberta}, and characterBERT\cite{ma2020charbert}) and three traditional deep learning models (CNN-LSTM\cite{ozcan2023hybrid}, BiLSTM\cite{huang2015bidirectional}, and TextCNN\cite{kim2014convolutional}) as baselines. Among these, BERT uses only MLM (Masked Language Modeling) for pre-training, without the next-sentence prediction task, since our dataset is not suited for that task. RoBERTa also employs MLM for pre-training but incorporates additional data augmentation techniques. CharacterBERT, on the other hand, uses MLM while operating at the character-level tokenization. The CNN-LSTM model combines a convolutional neural network (CNN) with a long short-term memory (LSTM) recurrent neural network. BiLSTM is a bidirectional LSTM, capturing dependencies in both forward and backward directions within the sequence. TextCNN, a purely CNN-based model, uses a text-specific method to initialize URL embeddings.

\begin{table}[t]
\caption{Summary of Static Feature Analysis; Underline indicates best performance within static feature-based and baseline methods respectively}
\label{tab:combined1}
\renewcommand{\arraystretch}{1.3}
\resizebox{\textwidth}{!}{
\begin{tabular}{|l|cllc|cllc|cllc|}
\hline
Methods: Static & \multicolumn{4}{c|}{\textbf{Phishing URL Detection}} & \multicolumn{4}{c|}{\textbf{Advertising URL Detection}} & \multicolumn{4}{c|}{\textbf{Webpage Classification}} \\ \cline{2-13} 
                                & Accuracy        & Recall          & F1-Score        & AUC             & Accuracy        & Recall          & F1-Score        & AUC             & Accuracy        & Recall          & F1-Score        & AUC             \\ \hline
urlBERT-KNN                     & 0.9587          & 0.9538          & 0.9414          & 0.9772          & 0.9728          & 0.9530          & 0.9489          & 0.9785          & 0.7112          & 0.6878          & 0.6839          & 0.7665          \\
urlBERT-LR                      & 0.9506          & 0.9389          & 0.9296          & 0.9743          & 0.9692          & 0.9536          & 0.9426          & 0.9744          & 0.7066          & 0.6711          & 0.6751          & 0.7810          \\
urlBERT-NB                      & 0.9627          & 0.9521          & 0.9466          & 0.9798          & 0.9706          & 0.9501          & 0.9448          & 0.9762          & 0.7199          & 0.6930          & 0.6921          & 0.7866          \\
urlBERT-RF                      & 0.9655          & 0.9503          & 0.9504          & 0.9829          & \underline{0.9866} & \underline{0.9638} & \underline{0.9744} & 0.9828          & 0.7233          & 0.7078          & 0.6992          & 0.7980          \\
urlBERT-GBM                     & 0.9631          & \underline{0.9577} & 0.9475          & 0.9714          & 0.9813          & 0.9590          & 0.9645          & 0.9865          & 0.7365          & 0.7058          & 0.7088          & 0.8015          \\
urlBERT-CNN                     & \underline{0.9688} & 0.9531          & \underline{0.9550} & 0.9833          & 0.9858          & 0.9676          & 0.9731          & \underline{0.9935} & \underline{0.7379} & \underline{0.7095} & \underline{0.7110} & \underline{0.8021} \\ \hline
\end{tabular}}
\end{table}

\begin{table}[t]
\caption{Summary of Baseline Comparison Experimental Results ( Underline indicates best performance within baseline methods respectively)}
\label{tab:combined2}
\renewcommand{\arraystretch}{1.3}
\resizebox{\textwidth}{!}{
\begin{tabular}{|l|cllc|cllc|cllc|}
\hline
Baseline/Our & \multicolumn{4}{c|}{\textbf{Phishing URL Detection}} & \multicolumn{4}{c|}{\textbf{Advertising URL Detection}} & \multicolumn{4}{c|}{\textbf{Webpage Classification}} \\ \cline{2-13} 
                                & Accuracy        & Recall          & F1-Score        & AUC             & Accuracy        & Recall          & F1-Score        & AUC             & Accuracy        & Recall          & F1-Score        & AUC             \\ \hline
characterBERT                           & 0.9567          & 0.9507          & 0.9385          & 0.9782          & 0.9828          & 0.9680          & 0.9676          & 0.9855          & 0.7012          & 0.6678          & 0.6859          & 0.7036          \\
RoBERTa                         & 0.9676          & \underline{0.9689} & \underline{0.9541} & \underline{0.9888} & \underline{0.9882} & 0.9716          & \underline{0.9776} & \underline{0.9844} & 0.6566          & 0.6811          & 0.6499          & 0.6880          \\
BERT                            & 0.9477          & 0.9421          & 0.9260          & 0.9759          & 0.9856          & 0.9621          & 0.9725          & 0.9833          & 0.6899          & 0.6730          & 0.6712          & 0.7256          \\
CNN-LSTM                        & 0.9565          & 0.9333          & 0.9372          & 0.9709          & 0.9796          & 0.9618          & 0.9615          & 0.9809          & 0.6263          & 0.6338          & 0.6077          & 0.6548          \\
BiLSTM                          & 0.9545          & 0.9437          & 0.9351          & 0.9784          & 0.9776          & 0.9550          & 0.9576          & 0.9805          & 0.6031          & 0.5893          & 0.5219          & 0.5755          \\
TextCNN                         & \underline{0.9637} & 0.9621          & 0.9485          & 0.9863          & 0.9820          & 0.9611          & 0.9659          & 0.9824          & \underline{0.7213} & \underline{0.7015} & \underline{0.7035} & \underline{0.7407} \\ \hline
\textbf{urlBERT}                & 0.9720          & 0.9691          & 0.9601          & 0.9921          & 0.9896          & 0.9721          & 0.9802          & 0.9935          & 0.7436          & 0.7360 & 0.7229          & 0.8111          \\ \hline
\end{tabular}}
\end{table}

\begin{table}[t]
\caption{Comparison of urlBERT and urlBERT-E on Different Tasks}
\label{tab:urlBERT_comparison}
\renewcommand{\arraystretch}{1.3}
\resizebox{0.5\textwidth}{!}{
\begin{tabular}{|l|c|c|c|c|c|}
\hline
\multirow{2}{*}{\textbf{Methods}} & \multicolumn{5}{c|}{\textbf{Phishing URL Detection}} \\ \cline{2-6}
                                 & Accuracy  & Recall   & F1-Score & AUC    & Time (min)  \\ \hline
urlBERT                          & 0.9720    & 0.9691   & 0.9601   & 0.9921 & 65.15    \\ 
urlBERTc                        & \textbf{0.9739} & \textbf{0.9699} & \textbf{0.9627} & \textbf{0.9932} & 28.27    \\ \hline

\multirow{2}{*}{\textbf{Methods}} & \multicolumn{5}{c|}{\textbf{Advertising URL Detection}} \\ \cline{2-6}
                                 & Accuracy  & Recall   & F1-Score & AUC    & Time (s)  \\ \hline
urlBERT                          & 0.9896    & 0.9721   & 0.9802   & 0.9935 & 25.42    \\ 
urlBERTc                        & \textbf{0.9935} & \textbf{0.9755} & \textbf{0.9876} & \textbf{0.9949} & 10.87    \\ \hline

\multirow{2}{*}{\textbf{Methods}} & \multicolumn{5}{c|}{\textbf{Webpage Classification}} \\ \cline{2-6}
                                 & Accuracy  & Recall   & F1-Score & AUC    & Time (s)  \\ \hline
urlBERT                          & 0.7436    & 0.7360   & 0.7229   & 0.8111 & 38.33    \\ 
urlBERTc                        & \textbf{0.7448} & 0.7293   & \textbf{0.7269} & \textbf{0.8206} & 13.89    \\ \hline
\end{tabular}}
\end{table}

\subsection{Performance of {\scshape urlBERT} as a Static URL Feature Extractor}
In this experiment, we evaluate the effectiveness of \textsc{urlBERT} as a static feature extractor. This approach involves freezing the parameters of the \textsc{urlBERT} encoder and using it directly to extract URL embedding features, which are then fed into various classifiers for training. The classifiers employed include K-Nearest Neighbors (KNN), Random Forest (RF), Convolutional Neural Network (CNN), Logistic Regression (LR), Naive Bayes (NB), and Gradient Boosting Machine (GBM). 

For the phishing detection and advertisement detection datasets, we employ a 60\% training and 40\% testing split. For the data set of web topic classification, we utilize a 80\% training and 20\% testing split. Given that the webpage topic classification task is defined as a multi-class problem, we adopt micro-averaging for the evaluation. Table \ref{tab:combined1} presents the results of using {\scshape urlBERT} as a static feature extractor for training various classifiers.
The experimental results demonstrate that the static features extracted by {\scshape urlBERT} are highly effective across various classifiers. Even traditional classifiers such as KNN, RF, and NB achieve over 96\% accuracy, with AUC scores approaching or exceeding 0.98. When CNN is used as the classifier, its performance is best. This indicates that {\scshape urlBERT}, after our pre-training process, can effectively capture the essential characteristics of diverse URLs, thus serving as a powerful feature extraction tool.

\subsection{Comparison with Other method}
We compare {\scshape urlBERT} with classical pre-trained models and deep learning methods, including BERT \cite{devlin2018bert}, RoBERTa \cite{liu2019roberta}, and characterBERT \cite{ma2020charbert}, as well as BiLSTM \cite{huang2015bidirectional}, CNN-LSTM \cite{ozcan2023hybrid}, and TextCNN \cite{kim2014convolutional}.
To ensure a fair comparison, all models were trained using the AdamW optimizer and consistent training and testing datasets. Learning rates were optimally tuned for each model architecture: 0.005 for BiLSTM, 2e-3 for TextCNN and TextRNN, and 2e-5 for {\scshape urlBERT}, BERT, RoBERTa, and XLNet. The batch size was standardized to 5 across all models, and testing was conducted after 5 training epochs.

Table \ref{tab:combined2} presents the comparative results of our proposed model and six baseline methods across three downstream tasks. {\scshape urlBERT} demonstrates consistent performance advantages over the baselines across different tasks. It achieves an AUC exceeding 0.99 in both phishing detection and advertisement detection tasks. For webpage topic classification, it attains an average AUC of 0.8111, surpassing the best baseline by 7\% and outperforming five other baselines (excluding TextCNN) by 10\%.
Observing the baseline methods, we note that their performance varies across different tasks (RoBERTa performs best in phishing detection and advertisement detection tasks, while TextCNN excels in webpage topic classification). In contrast, {\scshape urlBERT} displays a performance stability not exhibited by other methods.

\subsection{Evaluation of Improved Fine-Tuning}
We compare the performance of the standard fine-tuning approach with our two-stage fine-tuning method. The experimental results are presented in Table \ref{tab:urlBERT_comparison}. To differentiate the methods clearly, the two-stage fine-tuning approach is referred to as {\scshape urlBERTc}. The results indicate that {\scshape urlBERTc} offers improvements in the detection of phishing and advertisement URLs. Furthermore, the fine-tuning time of {\scshape urlBERTc} is substantially reduced compared to the standard fine-tuning approach. This efficiency gain can be attributed to the fact that, in the first stage, we refrain from fine-tuning the encoder, which is typically the most time- and resource-intensive component of the process.
\begin{figure}[t]
    \centering
    \includegraphics[width=0.5\linewidth]{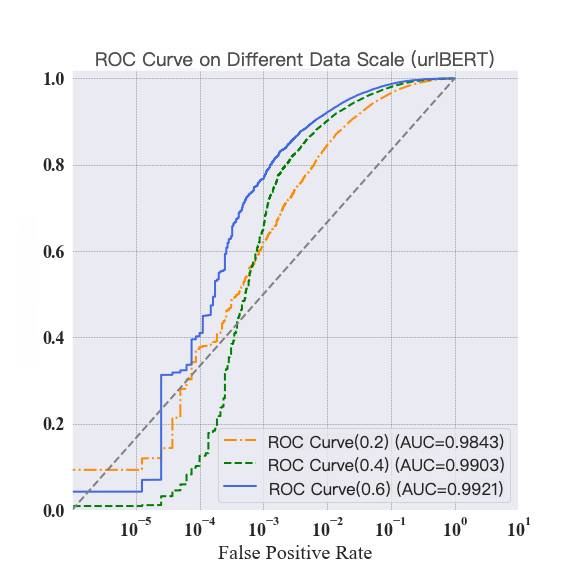}
    \caption{\textbf{ROC curves for {\scshape urlBERT} trained on training sets of different sizes.}}
    \label{fig:roc}
\end{figure}
\subsection{Evaluating the Impact of Data Scale on {\scshape urlBERT}}
This experiment assesses the model's sensitivity to training data scale, focusing solely on phishing URL detection for simplicity. We reduced the training data from 60\% to 5\%. BERT and RoBERTa, both pre-trained models, were used for comparison with our method. All experiments were conducted with consistent training settings over five epochs, using the same parameters as in previous evaluations.

\begin{table}[]
\caption{Phishing URL Detection Across Different Data Scales}
\label{tab:my-table}
\renewcommand{\arraystretch}{1.3}
\begin{tabular}{|c|c|c|c|c|c|c|}
\hline
\textbf{Scale} & \textbf{Models} & \textbf{Accuracy} & \textbf{Precision} & \textbf{Recall} & \textbf{F1-Score} & \textbf{AUC} \\ \hline
\multirow{3}{*}{0.6} & BERT & 0.9477 & 0.9003 & 0.9421 & 0.9260 & 0.9759 \\ \cline{2-7}
& RoBERTa  & 0.9637 & 0.9432 & 0.9621 & 0.9485 & 0.9863 \\ \cline{2-7}
& urlBERT & 0.9720 & 0.9513 & 0.9691 & 0.9601 & 0.9921 \\ \hline
\multirow{3}{*}{0.4} & BERT & 0.9401 & 0.9061 & 0.9233 & 0.9146 & 0.9678 \\ \cline{2-7}
& RoBERTa  & 0.9571 & 0.9427 & 0.9333 & 0.9380 & 0.9897 \\ \cline{2-7}
& urlBERT & 0.9661 & 0.9406 & 0.9633 & 0.9518 & 0.9903 \\ \hline
\multirow{3}{*}{0.2} & BERT & 0.9205 & 0.8676 & 0.9102 & 0.8884 & 0.9521 \\ \cline{2-7}
& RoBERTa  & 0.9505 & 0.9363 & 0.9202 & 0.9282 & 0.9757 \\ \cline{2-7}
& urlBERT & 0.9605 & 0.9388 & 0.9482 & 0.9435 & 0.9843 \\ \hline
\multirow{3}{*}{0.1} & BERT & 0.9156 & 0.8598 & 0.9047 & 0.8817 & 0.9321 \\ \cline{2-7}
& RoBERTa & 0.9206 & 0.8652 & 0.9140 & 0.8889 & 0.9435 \\ \cline{2-7}
& urlBERT & 0.9506 & 0.9164 & 0.9440 & 0.9300 & 0.9752 \\ \hline
\multirow{3}{*}{0.05} & BERT & 0.8721 & 0.7944 & 0.8527 & 0.8225 & 0.9069 \\ \cline{2-7}
& RoBERTa & 0.8864 & 0.8261 & 0.8527 & 0.8392 & 0.9238 \\ \cline{2-7}
& urlBERT & 0.9264 & 0.9164 & 0.9027 & 0.8875 & 0.9676 \\ \hline
\end{tabular}
\end{table}

The experimental results presented in Table \ref{tab:my-table} demonstrate the superiority of {\scshape urlBERT} in phishing URL detection across different data scales. {\scshape urlBERT} consistently outperforms both BERT and RoBERTa in terms of accuracy, precision, recall, F1-score, and AUC. At a 60\% data scale, it achieves the highest accuracy of 0.9720, compared to 0.9477 for BERT and 0.9637 for RoBERTa. Even at lower data scales, such as 10\%, {\scshape urlBERT} maintains robust performance, with an accuracy of 0.9506, surpassing both BERT (0.9156) and RoBERTa (0.9206).  These results highlight that {\scshape urlBERT} is particularly effective in handling reduced data, making it a strong and reliable choice for phishing URL detection in real-world scenarios. We also plotted the ROC curves and AUC values of {\scshape urlBERT} with varying training data scales of 0.6, 0.4, and 0.2, as shown in Figure \ref{fig:roc}. From the figure, it is evident that the model trained with 60\% of the training samples can still achieve a TPR close to 0.8 at a fixed FPR of 0.001.

\begin{table}[t]
\centering
\caption{Performance Comparison of Different Layer Configurations and Pooling Strategies}
\label{tab:layer-performance}
\renewcommand{\arraystretch}{1.2}
\resizebox{0.8\textwidth}{!}{%
\begin{tabular}{|l|cc|cc|cc|}
\hline
\multirow{2}{*}{\textbf{Layer Configuration}} & \multicolumn{2}{c|}{\textbf{Phishing Detection}} & \multicolumn{2}{c|}{\textbf{Advertising Detection}} & \multicolumn{2}{c|}{\textbf{Webpage Classification}} \\
\cline{2-7}
 & \textbf{Accuracy} & \textbf{AUC} & \textbf{Accuracy} & \textbf{AUC} & \textbf{Accuracy} & \textbf{AUC} \\
\hline
Layer-8 & 0.9713 & 0.9890 & 0.9891 & 0.9930 & 0.7389 & 0.7903 \\
Layer-9 & 0.9706 & 0.9898 & 0.9872 & 0.9921 & 0.7338 & 0.7890 \\
Layer-10 & \textbf{0.9717} & 0.9919 & 0.9888 & 0.9926 & 0.7410 & \textbf{0.8088} \\
Layer-11 & 0.9716 & 0.9919 & 0.9893 & \textbf{0.9939} & 0.7322 & 0.7907 \\
\hline
Last 4 Layers + Concat & 0.9713 & 0.9911 & 0.9890 & 0.9927 & 0.7299 & 0.7544 \\
Last 4 Layers + MeanPooling & 0.9707 & 0.9903 & 0.9877 & 0.9906 & 0.7066 & 0.7566 \\
Last 4 Layers + MaxPooling & 0.9612 & 0.9788 & 0.9834 & 0.9888 & 0.7013 & 0.7549 \\
Last 4 Layers + MinPooling & 0.9716 & 0.9896 & 0.9865 & 0.9921 & 0.7221 & 0.7854 \\
Last 4 Layers + WeightedPooling & 0.9718 & 0.9904 & 0.9888 & 0.9911 & \textbf{0.7440} & 0.8111 \\
Last 4 Layers + AttentionPooling & \textbf{0.9720} & \textbf{0.9921} & \textbf{0.9896} & 0.9935 & 0.7436 & 0.8066 \\
\hline
\end{tabular}%
}
\end{table}

\subsection{Multi-Task Evaluation}
URL-based tasks often share common characteristics, presenting an opportunity to develop unified multi-task models for practical applications. This training approach allows the execution and management of multiple URL-related tasks following a single supervised training session. In this experiment, we continued with supervised multitask learning during the fine-tuning phase, as described in Section \ref{MT}, and evaluated its performance differences between three tasks compared to fine-tuned models of a single task. We implement the multi-task learning architecture of this pre-trained model by following the multi-task learning setup inspired by the MT-DNN architecture \cite{liu2019multi}.To differentiate, we designate the multitask learning version as {\scshape urlBERT-MT}.

\begin{table*}[t]
    \centering
    \caption{Model Performance: Single-Task vs. Multi-Task Learning with {\scshape urlBERT.}}
    \label{tab:tab5}
    \renewcommand{\arraystretch}{1.2}
    \begin{tabular}{@{}lcccccc@{}}
        \toprule
        \textbf{Methods} & \multicolumn{2}{c}{\textbf{Phishing URL Detection}} & \multicolumn{2}{c}{\textbf{Advertising URL Detection}} & \multicolumn{2}{c}{\textbf{Webpage Classification}} \\
        \cmidrule(r){2-3} \cmidrule(lr){4-5} \cmidrule(l){6-7}
        & Accuracy & AUC & Accuracy & AUC & Accuracy & AUC \\
        \midrule
        {\scshape urlBERT}            & 0.9716 & 0.9943 & 0.9990 & 0.9995 & 0.7436 & 0.7986 \\
        {\scshape urlBERT-MT} & 0.9682 & 0.9913 & 0.9981 & 0.9984 & 0.7304 & 0.8101 \\
        \bottomrule
    \end{tabular}
\end{table*}

The experimental results presented in Table \ref{tab:tab5} show that multi-task learning (MTL) with {\scshape urlBERT-MT} performs comparably to the single-task version of {\scshape urlBERT} across all tasks. In phishing URL detection, {\scshape urlBERT} achieves an accuracy of 0.9716 and AUC of 0.9943, while {\scshape urlBERT-MT} obtains an accuracy of 0.9682 and AUC of 0.9913, indicating a minor performance difference. Similarly, for advertising URL detection, the accuracy and AUC values for {\scshape urlBERT-MT} (0.9981 and 0.9984) are only slightly lower than those for {\scshape urlBERT} (0.9990 and 0.9995). In webpage classification, although {\scshape urlBERT-MT} shows a small drop in accuracy (0.7304 vs. 0.7436), it compensates with a higher AUC of 0.8101 compared to 0.7986 for {\scshape urlBERT}. These results suggest that multi-task learning can achieve performance nearly on par with single-task learning, making it a viable approach for tasks that benefit from shared learning.

\subsection{Evaluating Model Efficacy across Training Epochs}
\begin{figure*}[t]
    \centering
    \includegraphics[width=0.9\linewidth]{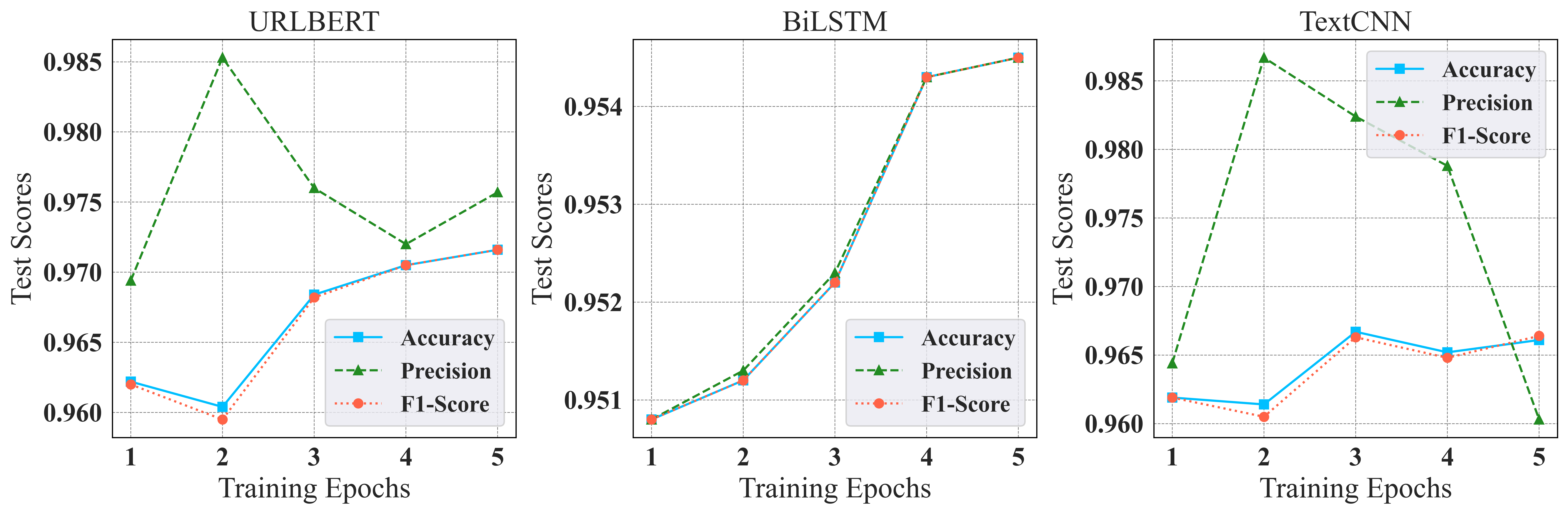}
    \caption{\textbf{Performance comparison across training epochs for {\scshape urlBERT}, BiLSTM, and TextCNN. Metrics include Accuracy, Precision, and F1-Score. All models are evaluated on the phishing URL detection task.}}
    \label{fig:Phish}
\end{figure*}

Figure \ref{fig:Phish} offers a detailed comparative analysis of core performance metrics—accuracy, precision, and F1-score—for {\scshape urlBERT}, BiLSTM, and TextCNN models across incremental training epochs during the phishing URL detection task. Notably, {\scshape urlBERT} establishes consistent high performance early in the training phase, peaking in precision during the second epoch, which underscores its capacity to achieve robust classification with fewer computational cycles. In contrast, the BiLSTM model shows a steady increase in all performance metrics over the course of training, indicating that its learning capability might be more data-dependent, requiring additional epochs to reach optimal performance levels. TextCNN displays a contrasting trend, with accuracy and F1-score peaking at the second epoch and then demonstrating a marked decline. This could point to a possible overfitting scenario where the model’s performance on training data is not generalizing well to the test set beyond a certain point, necessitating early stopping or regularization techniques to maintain peak performance.

\subsection{Transformer Layer Feature Extraction Proficiency}
This ablation study investigates the feature extraction capabilities of different layers within {\scshape urlBERT}. We extract [CLS] token representations from various layers during the fine-tuning process and evaluate their effectiveness when fed into the top-layer network for downstream task learning. In addition, we analyze the impact of different pooling strategies on downstream task performance. Throughout the experiments, we maintain consistent training parameters, employing the AdamW optimizer with a learning rate of 2e-5 and weight decay of 1e-4 over five training epochs.

Based on the experimental results presented in Table \ref{tab:layer-performance}, we can draw several insights into the performance of different layer configurations and pooling strategies across three distinct URL-related tasks. Among single layer configurations, Layer-10 and Layer-11 generally outperform the others, indicating that higher-level features captured in these layers are particularly relevant for URL-related tasks. This observation aligns with previous findings in NLP tasks where mid to high-level layers often contain the most task-relevant information.
Interestingly, strategies that combine the last 4 layers consistently outperform single layer configurations across all tasks. This suggests that integrating information from multiple layers provides a richer representation, capturing both low-level and high-level features crucial for URL analysis.
Among the pooling strategies, AttentionPooling demonstrates superior performance, especially in phishing detection where it achieves the highest accuracy (0.9720) and AUC (0.9921). This underscores the effectiveness of attention mechanisms in weighing the importance of different features for classification tasks. However, it's worth noting that WeightedPooling shows the best performance in webpage classification (accuracy 0.7440, AUC 0.8111), indicating that different tasks may benefit from tailored pooling strategies.

\section{Discussion} \label{section 6}

We propose {\scshape urlBERT}, an unsupervised URL Transformer encoder, and demonstrate its improvement over previous methods in both single-task and multi-task settings. The discussion focuses on its robustness and usability, scalability for downstream tasks, Model Efficiency, and future directions.

\begin{enumerate}
\item Robustness and Usability: Our experiments show that {\scshape urlBERT} consistently outperforms other pre-trained models and traditional deep learning methods across various tasks. It demonstrates higher robustness and flexibility, maintaining performance even with smaller datasets. Unlike previous URL classification models that required complex feature engineering and extensive parameter tuning, {\scshape urlBERT} offers a highly automated and user-friendly approach, eliminating the need for manual feature extraction and fine-tuning.
  
\item Scalability for Downstream Tasks: Following the "pre-train and fine-tune" paradigm, {\scshape urlBERT} ensures excellent scalability across different URL tasks. While we focused on three URL classification tasks, it can easily be applied to a wide range of internet-related applications, such as spam detection, content moderation, social media link analysis, and news aggregation.

\item Model Efficiency: While \textsc{urlBERT} demonstrates strong performance, its Transformer-based architecture incurs significant computational and memory overhead, which can limit real-time deployment on resource-constrained environments such as mobile devices or browser-based security plugins. To address this, future work will explore model compression techniques including knowledge distillation, quantization, pruning, and low-rank factorization to reduce inference latency and memory footprint. These approaches have proven effective in compressing large language models while retaining much of their performance, and we anticipate similar benefits when applied to \textsc{urlBERT}. By leveraging such strategies, it is feasible to develop lightweight variants of \textsc{urlBERT} suitable for edge deployment and real-time malicious URL detection.

\item Future Directions: Despite comprehensive evaluation, there are areas for further research. Class-incremental learning (CIL) could allow {\scshape urlBERT} to incrementally incorporate knowledge of new URL categories, building a general-purpose classifier. This approach is critical for adapting to the diverse range of URL classification tasks encountered on the internet. Additionally, few-shot or zero-shot learning, enabled by pre-trained models, is an important avenue for future work, enabling rapid adaptation to new tasks with minimal data.
\end{enumerate}

\section{Conclusion}\label{section 7}
Applying pre-trained models based on the Transformer architecture to various downstream tasks for improving model performance has become a widely adopted practice in fields such as natural language processing, biomedical corpora, code language generation, and more. In this work, inspired by this approach, we have employed carefully designed pre-training tasks to pre-train a model, {\scshape urlBERT}, based on BERT on a large amount of unlabeled URL data, aiming to facilitate URL analysis in research. Our training tasks enable the model to fully grasp both structural and semantic features of URLs, allowing the foundational BERT model to adapt comprehensively to the semantic characteristics of URL text. Through well-designed experiments in diverse scenarios, the effectiveness of {\scshape urlBERT} has been validated. Consequently, when facing various downstream tasks in the field of URL analysis, {\scshape urlBERT} can be efficiently fine-tuned as a feature encoder for deployment in specific tasks.

\bibliographystyle{splncs04}
\bibliography{bibtex/bib/ref}

\begin{thebibliography}{10}
\providecommand{\url}[1]{\texttt{#1}}
\providecommand{\urlprefix}{URL }
\providecommand{\doi}[1]{https://doi.org/#1}

\bibitem{ali2024efficient}
Ali~Reshi, J., Ali, R.: An efficient fake news detection system using contextualized embeddings and recurrent neural network  (2024)

\bibitem{artene2021using}
Artene, C.G., Tibeic{\u{a}}, M.N., Leon, F.: Using bert for multi-label multi-language web page classification. In: 2021 IEEE 17th International Conference on Intelligent Computer Communication and Processing (ICCP). pp. 307--312. IEEE (2021)

\bibitem{baykan2009purely}
Baykan, E., Henzinger, M., Marian, L., Weber, I.: Purely url-based topic classification. In: Proceedings of the 18th international conference on World wide web. pp. 1109--1110 (2009)

\bibitem{grambeddings}
Bozkir, A.S., Dalgic, F.C., Aydos, M.: Grambeddings: a new neural network for url based identification of phishing web pages through n-gram embeddings. Computers \& Security  \textbf{124},  102964 (2023)

\bibitem{devlin2018bert}
Devlin, J., Chang, M.W., Lee, K., Toutanova, K.: Bert: Pre-training of deep bidirectional transformers for language understanding. arXiv preprint arXiv:1810.04805  (2018)

\bibitem{do2022deep}
Do, N.Q., Selamat, A., Krejcar, O., Herrera-Viedma, E., Fujita, H.: Deep learning for phishing detection: Taxonomy, current challenges and future directions. Ieee Access  \textbf{10},  36429--36463 (2022)

\bibitem{elsadig2022intelligent}
Elsadig, M., Ibrahim, A.O., Basheer, S., Alohali, M.A., Alshunaifi, S., Alqahtani, H., Alharbi, N., Nagmeldin, W.: Intelligent deep machine learning cyber phishing url detection based on bert features extraction. Electronics  \textbf{11}(22), ~3647 (2022)

\bibitem{fernandez2018learning}
Fern{\'a}ndez, A., Garc{\'\i}a, S., Galar, M., Prati, R.C., Krawczyk, B., Herrera, F.: Learning from imbalanced data sets, vol.~10. Springer (2018)

\bibitem{gao2021simcse}
Gao, T., Yao, X., Chen, D.: Simcse: Simple contrastive learning of sentence embeddings. arXiv preprint arXiv:2104.08821  (2021)

\bibitem{garcia2018dynamic}
Garc{\'\i}a, S., Zhang, Z.L., Altalhi, A., Alshomrani, S., Herrera, F.: Dynamic ensemble selection for multi-class imbalanced datasets. Information Sciences  \textbf{445},  22--37 (2018)

\bibitem{goodfellow2014explaining}
Goodfellow, I.J., Shlens, J., Szegedy, C.: Explaining and harnessing adversarial examples. arXiv preprint arXiv:1412.6572  (2014)

\bibitem{haynes2021lightweight}
Haynes, K., Shirazi, H., Ray, I.: Lightweight url-based phishing detection using natural language processing transformers for mobile devices. Procedia Computer Science  \textbf{191},  127--134 (2021)

\bibitem{he2023method}
He, D., Lv, X., Zhu, S., Chan, S., Choo, K.K.R.: A method for detecting phishing websites based on tiny-bert stacking. IEEE Internet of Things Journal  (2023)

\bibitem{huang2015bidirectional}
Huang, Z., Xu, W., Yu, K.: Bidirectional lstm-crf models for sequence tagging. arXiv preprint arXiv:1508.01991  (2015)

\bibitem{joulin2016bag}
Joulin, A., Grave, E., Bojanowski, P., Mikolov, T.: Bag of tricks for efficient text classification. arXiv preprint arXiv:1607.01759  (2016)

\bibitem{kim2014convolutional}
Kim, Y.: Convolutional neural networks for sentence classification (2014)

\bibitem{urlnet}
Le, H., Pham, Q., Sahoo, D., Hoi, S.C.: Urlnet: Learning a url representation with deep learning for malicious url detection. arXiv preprint arXiv:1802.03162  (2018)

\bibitem{liu2024transurl}
Liu, R., Wang, Y., Guo, Z., Xu, H., Qin, Z., Ma, W., Zhang, F.: Transurl: Improving malicious url detection with multi-layer transformer encoding and multi-scale pyramid features. Computer Networks  \textbf{253},  110707 (2024)

\bibitem{liu2023malicious}
Liu, R., Wang, Y., Xu, H., Qin, Z., Liu, Y., Cao, Z.: Malicious url detection via pretrained language model guided multi-level feature attention network. arXiv preprint arXiv:2311.12372  (2023)

\bibitem{liu2025pmanet}
Liu, R., Wang, Y., Xu, H., Qin, Z., Zhang, F., Liu, Y., Cao, Z.: Pmanet: Malicious url detection via post-trained language model guided multi-level feature attention network. Information Fusion  \textbf{113},  102638 (2025)

\bibitem{liu2019multi}
Liu, X., He, P., Chen, W., Gao, J.: Multi-task deep neural networks for natural language understanding. arXiv preprint arXiv:1901.11504  (2019)

\bibitem{liu2019roberta}
Liu, Y., Ott, M., Goyal, N., Du, J., Joshi, M., Chen, D., Levy, O., Lewis, M., Zettlemoyer, L., Stoyanov, V.: Roberta: A robustly optimized bert pretraining approach. arXiv preprint arXiv:1907.11692  (2019)

\bibitem{lopez2013insight}
L{\'o}pez, V., Fern{\'a}ndez, A., Garc{\'\i}a, S., Palade, V., Herrera, F.: An insight into classification with imbalanced data: Empirical results and current trends on using data intrinsic characteristics. Information sciences  \textbf{250},  113--141 (2013)

\bibitem{ma2020charbert}
Ma, W., Cui, Y., Si, C., Liu, T., Wang, S., Hu, G.: Charbert: Character-aware pre-trained language model. arXiv preprint arXiv:2011.01513  (2020)

\bibitem{mikolov2013distributed}
Mikolov, T., Sutskever, I., Chen, K., Corrado, G.S., Dean, J.: Distributed representations of words and phrases and their compositionality. Advances in neural information processing systems  \textbf{26} (2013)

\bibitem{dmoz1}
Nokhbeh~Zaeem, R., Barber, K.S.: A large publicly available corpus of website privacy policies based on dmoz. In: Proceedings of the Eleventh ACM Conference on Data and Application Security and Privacy. pp. 143--148 (2021)

\bibitem{opara2024look}
Opara, C., Chen, Y., Wei, B.: Look before you leap: Detecting phishing web pages by exploiting raw url and html characteristics. Expert Systems with Applications  \textbf{236},  121183 (2024)

\bibitem{otieno2023detecting}
Otieno, D.O., Abri, F., Namin, A.S., Jones, K.S.: Detecting phishing urls using the bert transformer model. In: 2023 IEEE International Conference on Big Data (BigData). pp. 2483--2492. IEEE (2023)

\bibitem{ozcan2023hybrid}
Ozcan, A., Catal, C., Donmez, E., Senturk, B.: A hybrid dnn--lstm model for detecting phishing urls. Neural Computing and Applications pp. 1--17 (2023)

\bibitem{sahingoz2019machine}
Sahingoz, O.K., Buber, E., Demir, O., Diri, B.: Machine learning based phishing detection from urls. Expert Systems with Applications  \textbf{117},  345--357 (2019)

\bibitem{sahoo2017malicious}
Sahoo, D., Liu, C., Hoi, S.C.: Malicious url detection using machine learning: A survey. arXiv preprint arXiv:1701.07179  (2017)

\bibitem{seyyar2022detection}
Seyyar, Y.E., Yavuz, A.G., {\"U}nver, H.M.: Detection of web attacks using the bert model. In: 2022 30th Signal Processing and Communications Applications Conference (SIU). pp.~1--4. IEEE (2022)

\bibitem{Mendeley}
Singh, A.: Malicious and benign webpages dataset. Data in brief  \textbf{32},  106304 (2020)

\bibitem{srinivasan2021durld}
Srinivasan, S., Vinayakumar, R., Arunachalam, A., Alazab, M., Soman, K.: Durld: Malicious url detection using deep learning-based character level representations. Malware analysis using artificial intelligence and deep learning pp. 535--554 (2021)

\bibitem{su2023bert}
Su, M.Y., Su, K.L.: Bert-based approaches to identifying malicious urls. Sensors  \textbf{23}(20), ~8499 (2023)

\bibitem{sun2025ethereum}
Sun, J., Jia, Y., Wang, Y., Tian, Y., Zhang, S.: Ethereum fraud detection via joint transaction language model and graph representation learning. Information Fusion  \textbf{120},  103074 (2025)

\bibitem{sun2020ernie}
Sun, Y., Wang, S., Li, Y., Feng, S., Tian, H., Wu, H., Wang, H.: Ernie 2.0: A continual pre-training framework for language understanding. In: Proceedings of the AAAI conference on artificial intelligence. vol.~34, pp. 8968--8975 (2020)

\bibitem{tian2025past}
Tian, Y., Yu, Y., Sun, J., Wang, Y.: From past to present: A survey of malicious url detection techniques, datasets and code repositories. arXiv preprint arXiv:2504.16449  (2025)

\bibitem{tsai2024toward}
Tsai, Y.D., Liow, C., Siang, Y.S., Lin, S.D.: Toward more generalized malicious url detection models. In: Proceedings of the AAAI Conference on Artificial Intelligence. vol.~38, pp. 21628--21636 (2024)

\bibitem{wang2019tune}
Wang, R., Su, H., Wang, C., Ji, K., Ding, J.: To tune or not to tune? how about the best of both worlds? arXiv preprint arXiv:1907.05338  (2019)

\bibitem{wang2023large}
Wang, Y., Zhu, W., Xu, H., Qin, Z., Ren, K., Ma, W.: A large-scale pretrained deep model for phishing url detection. In: ICASSP 2023-2023 IEEE International Conference on Acoustics, Speech and Signal Processing (ICASSP). pp.~1--5. IEEE (2023)

\bibitem{wei2020accurate}
Wei, W., Ke, Q., Nowak, J., Korytkowski, M., Scherer, R., Wo{\'z}niak, M.: Accurate and fast url phishing detector: a convolutional neural network approach. Computer Networks  \textbf{178},  107275 (2020)

\bibitem{yan2021consert}
Yan, Y., Li, R., Wang, S., Zhang, F., Wu, W., Xu, W.: Consert: A contrastive framework for self-supervised sentence representation transfer. arXiv preprint arXiv:2105.11741  (2021)

\bibitem{zhang2021multi}
Zhang, W., Yang, G., Zhang, N., Xu, L., Wang, X., Zhang, Y., Zhang, H., Del~Ser, J., de~Albuquerque, V.H.C.: Multi-task learning with multi-view weighted fusion attention for artery-specific calcification analysis. Information Fusion  \textbf{71},  64--76 (2021)

\end{thebibliography}

\end{document}